\documentclass[10pt,conference]{IEEEtran}
\usepackage[utf8]{inputenc}
\usepackage[T1]{fontenc}
\usepackage{float}
\usepackage{amsmath}
\usepackage{amsfonts}
\usepackage{mathtools}
\usepackage{amsfonts}
\usepackage{amssymb}
\usepackage{makeidx}
\usepackage{csquotes}
\usepackage{graphicx}
\usepackage{cite}
\usepackage{xcolor}
\usepackage{subcaption}
\usepackage{graphicx}
\usepackage{balance}
\usepackage[capitalise,nameinlink ,noabbrev]{cleveref}
\usepackage[algo2e,ruled, vlined, linesnumbered]{algorithm2e}
\usepackage{ifpdf}
\ifpdf
\else
  
\fi
\usepackage[margin=1in]{geometry}
\begin{document}
\title{\huge{Multi-Objective Optimization for 3D Placement and Resource Allocation in OFDMA-based Multi-UAV Networks}\vspace{-4mm}
}

\author{Asad Mahmood, Thang X. Vu, Shree Krishna Sharma, Symeon Chatzinotas, Bj\"orn Ottersten \\Interdisciplinary Centre for Security, Reliability and Trust (SnT), University of Luxembourg\\ Email: \{asad.mahmood, thang.vu, symeon.chatzinotas, bjorn.ottersten\}@uni.lu; shree.sharma@ieee.org\vspace{-2mm}
}%
\maketitle
\vspace{-3mm}
\begin{abstract}
This work considers the orthogonal frequency division multiple access (OFDMA) technology that enables multiple unmanned aerial vehicles (multi-UAV) communication systems to provide on-demand services. The main aim of this work is to derive the optimal allocation of radio resources, $3D$ placement of UAVs, and user association matrices. To achieve the desired objectives, we decoupled the original joint optimization problem into two sub-problems: i) $3D$ placement and user association and ii) sum-rate maximization for optimal radio resource allocation, which are solved iteratively. The proposed iterative algorithm is shown via numerical results to achieve fast convergence speed after less than 10 iterations. The benefits of the proposed design are demonstrated via superior sum-rate performance compared to existing reference designs. Moreover, the results declared that the optimal power and sub-carrier allocation helped mitigate the co-cell interference that directly impacts the system's performance. 
\end{abstract}
\vspace{-1mm}
\section{Introduction}
\vspace{-0.5mm}
\subsection{Background and Related work}
\vspace{-1mm}
\IEEEPARstart{T}he $5$G and beyond communication systems aim to provide massive connectivity, ultra-reliable low latency communication, and ensure the end-user quality of service. In contrast, most of the earth's remote areas still need to be covered by the terrestrial base stations due to the sparse human activities in these regions \cite{7536859}. Moreover, the excessive demand for the users' resources and connectivity represents challenging tasks to meet the objectives mentioned earlier \cite{6824752,9001132,9580711}. In comparison, the UAV emerges as a practical solution because of its flexible deployment and adaptive altitude. Furthermore, the existing UAV cases will be more beneficial by integrating the UAVs with the base station, also known as a flying base station (FBS), to provide services to users located in remote areas.  
\par
Besides providing ubiquitous coverage, UAVs have several limitations, including security, privacy, flight regulations, and limited battery life  \cite{8660516}. The hovering time and services provided by UAVs are primarily determined by battery life. To overcome the limited battery capacity of the UAV, a highly energy-efficient mechanism for maintaining users' quality of service requirements is required \cite{7888557,8038869}. Similarly, optimal UAV placement is critical for ensuring a balanced trade-off between interference, shadowing, and path loss when multiple UAVs are used. Additionally, recent research indicates that elevating UAVs increases the probability of line-of-sight communication and introduces additional path loss. \cite{7536859,7417609}. As a result of the aforementioned challenges, UAVs must adapt their positions in response to changes in UE density to provide communication services as needed.
\par
Recent research has focused on optimizing UAV placement in 3D to maximize coverage while minimizing transmit power\cite{9044857,7486987,8744514,7918510,7841993,8821282}. Furthermore, it is possible to deploy multiple UAVs without interfering with each other while minimizing transmit power by optimizing the users association matrix, as described in \cite{9044857}. This work's primary goal is to optimize power and trajectory to maximize minimum secrecy. To assess the impact of altitude on interference, the authors of \cite{7486987} used multiple UAVs with directional antennas; the results show that adjusting the UAV's altitude helps mitigate the interference. It is also recommended that the distance between multiple UAVs be more significant than a threshold to avoid co-channel interference \cite{8744514}. To serve as many users as possible, the authors of \cite{7918510} propose a UAV placement algorithm that uses the least transmission power. The authors of \cite{7841993} proposed a method for reducing energy consumption by optimizing unmanned aerial vehicle trajectory and deployment.
\par
Aside from massive communication and ultra-reliable low latency communication, energy efficiency and resource efficiency are important performance metrics in $5G$ and beyond communication systems. For effective utilization of resources, the main aim of this work is to place the minimum number of UAVs that guarantee quality-of-service for all UEs in coverage area.  In a similar vein, the problem of minimizing the total number of UAVs has been studied recently \cite{9313991,7881122,7762053}. However, most of these works have considered fixed flying altitudes or UAVs as complementary parts of existing terrestrial communication systems during potential congestion periods. A closely related work is given in  \cite{8821282} in which the author considered the concept of simultaneous wireless information and power transfer (SWIPT), in which a multiple (fixed) number of UAVs are deployed by an iterative algorithm based on exhaustive search without considering the maximum serving capacity of each UAV and co-cell interference.
\vspace{-0.5mm}
\subsection{Contributions}
\vspace{-0.5mm}
In this work, we consider multi-UAV communication systems that use orthogonal frequency division multiple access (OFDMA) to provide heterogeneous services in areas where terrestrial base stations are unavailable due to natural disasters or UE congestion \cite{9348616}. We propose a novel objective that minimizes the number of deployed UAVs while maximizing the system sum rate via optimal radio resource allocation, user association matrix optimization, and efficient $3D$ placement of UAVs. To overcome the difficulty, we decoupled the original optimization problem into two sub-problems using block coordinate descent (BCD). For path loss minimization, the $3D$ placement and user association problems are solved using quadratic constraint quadratic programming (QCQP) and branch and bound (BnB) techniques, respectively. The sum-rate maximization problem is then solved using the successive convex approximation (SCA) technique. The performance of the proposed algorithm is demonstrated numerically, with rapid convergence occurring within ten iterations and superior sum-rate performance compared to the baseline scheme.
\par
The following sections summarize the rest of the paper: Section \ref{System Model and Problem Statement} summarizes the proposed system model and problem statement. Section \ref{Proposed Iterative Algorithm} presented an algorithm for efficient $3D$ UAV deployment and resource allocation. Likewise, Section \ref{Results and Discussion} contains the simulation results. The paper is finally concluded in Section \ref{Conclusion}.
\vspace{-1mm}
\section{System Model and Problem Statement}
\label{System Model and Problem Statement}
\vspace{-0.4mm}
This work considers the down-link UAV-assisted wireless communication system that provides heterogeneous services to users in areas where terrestrial base stations are unavailable due to natural disasters or UE congestion. The considered system model comprises $M$ number of available UAVs and $N$ number of users, represented by sets $\mathcal{M}\!\! \!= \!\!\!\{ 1,2, \cdots {\rm{M\} }}$ and $\mathcal{N}\!\! =\!\! \{ 1,2, \cdots {\rm{N\} }}$,  respectively. In order to improve the system's energy efficiency, only $L\!\leq\! M$ UAVs will be activated to serve the users such that $\mathcal{L}\! \subset\! \mathcal{M}$. We consider $3D\!$ coordinates, in which the location of $l$-th UAV and the $n$-th UE are denoted by $x^l,y^l,h^l$ and $x_n,y_n,0$, respectively. Both UAVs and UEs are equipped with single omnidirectional antennas because of their uniform radiation in all directions\cite{4067672}. The UAVs are assisted by a central controller that provides information about UE's location. To avoid mutual interference between connected UEs, each UAV serves a cluster of users using OFDMA. $Q_l\!$ represents users associated with the $l$-th UAV; each user is served by only one UAV. 
\par
In the considered system model, the frequency bandwidth is distributed among $K$ sub-channels donated by set $\mathcal{K}\!\!=\!\!\!\left\{ {1, \cdots ,K} \right\}$. We define $\textbf{A}=[a_n^{k,l}]_{N \times K \times L}$ as the sub-channel allocation  matrix, in which $a_n^{k,l}\!\!=\!\!1$ indicates that sub-channel $k$ allocated to user $n$ associated with UAV $l$, and $a_n^{k,l}\!\!=\!\!0$ otherwise. 
Additionally, we refer to $p_n^{k,l}$ as the transmit power transmitted from the UAV $l$ to the user $n$ via sub-channel $k$. Given that UAV $l$ consumes power on sub-channel $k$ only when user $n$ is assigned to that sub-channel, expressed mathematically as follows:
\vspace{-1mm}
\begin{equation}
 \label{Power allocation}
 \small
    p_n^{k,l} \!\le\! a_n^{k,l}P_t^{Max},
    \vspace{-2mm}
\end{equation}
Whereas $P_t^{Max}$ represents UAV's maximum transmits power. The new transmit power constraint \eqref{Power allocation} not only facilitates the rate function but also guarantees no transmit power on sub-channel $k$ unless user $k$ is assigned to it. To avoid cross-user interference, each sub-channel in a cluster is assigned to only one user, which imposes the following constraint.
\vspace{-1mm}
\begin{equation}
\small
    \sum\limits_{n \in \mathcal{N}} {{\rm{a}}_n^{k,l}}\!  \le\! 1,\forall k \in \mathcal{K},l \in \mathcal{L}.
    \vspace{-2mm}
\end{equation}
\vspace{-1mm}
\par
To model the UAVs placement, we introduce a binary vector $\Vec{I}\!\!=\!\![I_1,I_2,I_3 \cdots I_M]$, in which $\!I_m\!\!=\!1$ means the $m$-th UAV is activated, and $\!I_m\!\!=\!0$ otherwise. Similarly, the user association $\textbf{J}\!\!=\!\![{J_{n}^l}]{_{N \!\times L}}$ is a binary matrix, in which each entity $\!J_{n}^l$ represents the connection between $l$-th UAV and $n$-th UE, i.e., $\!J_n^l\!\!=\!1$ if user $n$ is associated with UAV $l$, and $\!J_n^l\!\!=\!0$ otherwise. Consequently, the following constraint must be met as one sub-channel is only allocated to a user if it is connected to the corresponding UAV:  
\vspace{-1mm}
\begin{equation}
\small
    \sum\limits_{k \in {\cal K}} {{\rm{a}}_n^{k,l}} \! \le\! J_n^l,\forall n \in {\cal N},l \in {\cal L}.
    \vspace{-1.5mm}
\end{equation}
\par
The quality wireless link between UAV-UE is measured in terms of the signal-to-interference plus noise ratio (SINR), which is expressed as follows:
\vspace{-1mm}
\begin{equation}
\small
 \label{SINR}
\gamma{_n^{k,l}}\! =\! \frac{{p_n^{k,l}g_n^{k,l}}}{{\Phi _n^{k,l}  + {\sigma ^2}}}.
\vspace{-1.5mm}
 \end{equation}
 In \eqref{SINR}, $\sigma\!$ is the Gaussian noise power, and $\Phi _n^{k,l}\!$ is the aggregated interference imposed by the UAVs on the $n$-th UE except the $l$-th UAV, which is represented as $\Phi _n^{k,l}\! =\!\sum_{l \ne l'} \!{\left( {\sum_{n \ne n'}p_{n'}^{k,l'}} \right)} g_n^{k,l'}$. The UAV-UE channel gain 
 $g_n^{k,l}$ is computed as  
  $   g _n^{k,l}\!\!\!=\!\!\!\frac{\left | h_n^{k,l} \right|^2}{{PL}_n^{l}}$,
 where $h_n^{k,l}\!\sim \!CN\! \left(0,1 \right )$ is the small-scale fading and $PL_n^l$ is the path-loss. In this work, we consider the air-to-ground (A2G) channel model that is composed of both line of sight (LoS) and non-line of sight (NLoS) components. Mathematically the path loss is expressed as follows \cite{8727504}:
\begin{equation}
\label{channel}
\small
{PL}_n^{l}\!=\!\!K_o\left(d_n^l\right)^\alpha \left[PLos_n^l\xi_{LoS}\!+\!{\xi_{NLoS}}PNLos_{n}^l\right],
\vspace{-1.5mm}
\end{equation}
where $K_o \!\!= \!\!{\left( {\frac{{4\pi {f_c}}}{c}} \right)^2}$ and $\xi_{LoS}$ and $\mu_{NLoS}$ are the attenuation factors  for LoS and NLoS link. Similarly, $f_c$, $\alpha$ and $c$ represent the carrier frequency, path loss exponent and speed of light, respectively. $PLos_n^l \!\!= \!\!\frac{1}{{b_1}\exp \left( { - {b_2}(\theta _n^l - {b_1})}\right)}$ and $PNLos_n^l\!\! = \!\!1\! -\! PLos_n^l$ represent the probability of the LoS and NLoS, respectively, where $b_1$ and $b_2$  are the constants representing the environment condition, and $\theta _{n}^l$ is the angle of elevation between the $l$-th UAV and the $n$-th UE. Thus, the achievable rate of $n$-th UE connected to $l$-th UAV is represented as follows:
\begin{equation}
\small
    \label{Rate carrier}
{R_n^{l}} = {B}{\sum}_{k \in \mathcal{K}}{\log _2}\left( {1 + \gamma{_n^{k,l}} } \right),
\vspace{-2mm}
\end{equation}
where $\gamma_n^{k,l}$ is given in \eqref{SINR}.
\vspace{-1mm}
\subsection{Problem Formulation}
\vspace{-0.5mm}
Unlike previous works, which target only the sum-rate maximization, we also optimize the system-wise energy usage. Therefore, this work aims to maximize the ratio between the system sum rate and the number of active UAVs, which is given as $\Upsilon\!\! =\!\! \frac{\sum_{n,l}R_n^l}{L}\!$, where $R_n^l\!$ is calculated in \eqref{Rate carrier} and $L\!$ is the number of active UAVs. We aim to maximize the objective function $\!\!\Upsilon\!\!$ by optimizing the UAVs' 3D placement, user association, power, and sub-carrier allocation. 
Before going to the mathematical problem formulation, we introduce designing variables. The 3D location of each UAV is denoted by $\textbf{W}\!\!=\!\! \left[ {{x^{l}},{y^{l}},{h^{l}}} \right]$; power profile, user association matrix and UAVs placement vector, represented as $\textbf{P}\!\!=\!\! [{p_{n}^{k,l}}]{_{N \times\! K \times L}}$, $\textbf{J}$ and $\Vec{I}$, respectively; whereas sub-channel allocation matrix is represented by $\textbf{A}$. Based on these compact variable notations, the joint optimization is formulated as follows:
\vspace{-1.5mm}
\begin{subequations}
\small
\vspace{-2mm}
	\label{Main Equation1}
	\begin{align}
		\textbf{P1}:\;\;\;&\mathop {\max }\limits_{\textbf{W},\textbf{P},\textbf{J},\Vec{I},\textbf{A}} \frac{\sum_{l\in\mathcal{L}}\sum_{n\in \mathcal{N}}R_n^l}{L}, \\[-1pt]
	\label{C1}	\textbf{C1}:\;\;\;&{R_n^l \ge J_{nl}R_n},\forall n\in \mathcal{N}, l\in \mathcal{L}, \\[-1pt]
	\label{C2}
	\textbf{C2}:\;\;\;&\sum\limits_{{l} \in \mathcal{L}} {J_{n}^l} \le 1 ,,\forall n \in \mathcal{N}, \\[-1pt]
	\label{C3}
	\textbf{C3}:\;\;\;&\sum\limits_{{n} \in \mathcal{N}} {{J_{n}^l} \le C^l_{Max}},\forall l \in \mathcal{L},  \\[-1pt]
	\label{C4}
	\textbf{C4}:\;\;\;&\sum\limits_{{n} \in \mathcal{N}} {{J_{n}^l} \ge C^l_{Min}} ,\forall l \in \mathcal{L}, \\[-1pt]
		\label{C5}
	\textbf{C5}:\;\;\;& {p_n^{k,l} \le a^{k,l}_n P_t^{Max}},\forall n \in \mathcal{N},k \in \mathcal{K},l \in \mathcal{L},  \\[-1pt]
		\label{C6}
	\textbf{C6}:\;\;\;&\sum\limits_{n \in \cal{N}} {\sum\limits_{k \in \mathcal{K}} {p_n^{k,l} \le P_t^{Max}} },\forall l \in \mathcal{L},  \\[-1pt]
	\label{C7}
	\textbf{C7}:\;\;\;&\sum\limits_{k \in {\cal K}} {{\rm{a}}_n^{k,l}}  \le J_n^l,\forall n \in {\cal N},l \in {\cal L}, \\[-1pt]
	\label{C8}
	\textbf{C8}:\;\;\;&\sum\limits_{n \in \mathcal{N}} {{\rm{a}}_n^{k,l}}  \le 1,\forall k \in \mathcal{K},l \in \mathcal{L}, \\[-1pt]
	\label{C9}
	\textbf{C9}:\;\;\;&\mathop {{\hbar _{(1 \times N)}}{\bf{J}}{\mathchar'26\mkern-10mu\lambda _{(L \times 1)}}}\limits_{\forall {\hbar _n},\forall {\mathchar'26\mkern-10mu\lambda _l} = 1}  \ge \lambda N, \\[-1pt]
		\label{C10} 
	\textbf{C10}:\;\;\;&\mathop {\left\| {{W^l} - {W^{l'}}} \right\|} \ge {d_{_o}},{\forall l' \in \mathcal{L}}, \\[-1pt]
	\label{C11}
	\textbf{C11}:\;\;\;&\{ \mathop {\left. {{x_l}} \right|{x_l} \in \mathbb{R}}:x_{min} \le {x_l} \le {x_{\max }}\},{\forall l \in \mathcal{L}} , \\[-1pt]
	\label{C12}
	\textbf{C12}:\;\;\;&\{ \mathop {\left. {{y_l}} \right|{y_l} \in \mathbb{R}}:y_{\min} \le {y_l} \le {y_{\max }}\},{\forall l \in \mathcal{L}}, \\[-1pt]
	\label{C13}
	\textbf{C13}:\;\;\;&\{ \mathop {\left. {{h_l}} \right|{h_l} \in \mathbb{R}}:h_{\min} \le {h_l} \le {h_{\max }}\},{\forall l \in \mathcal{L}}, \\[-1pt]
	\label{C14}
	\textbf{C14}:\;\;\;&\{ \mathop {\left. {{{p_n^{k,l}}}} \right|{{p_n^{k,l}}} \in \mathbb{R}}:{p_n^{k,l}} \ge 0\},  {\forall {p_n^{k,l}} \in \textbf{P}},\\[-1pt]
	\label{C15}
	\textbf{C15}:\;\;\;& I_m,\! J_n^l,\! a_n^{k,l}\! \in \{0,1\},\!\forall I_m,J_n^l,a_n^{k,l}\in \Vec{I},\textbf{J},\! \textbf{A}.
    \end{align}
\end{subequations}
The primary objective of this work is to deploy the minimum number of UAVs possible while maximizing the average rate per UAV, subject to certain user and system constraints detailed in $\textbf{C1}$ to $\textbf{C15}$. The quality of service constraint $\textbf{C1}$ requires that each UE's rate be above the minimum threshold. $\textbf{C2}$ prevents any UE from communicating with more than one UAV. Whereas $\textbf{C3}$ and $\textbf{C4}$ indicate the UAVs' maximum and maximum serving capacity, respectively. Similarly, the constraint $\textbf{C5}$ indicates the power allocation over the sub-carrier $k$. $\textbf{C6}$ specifies that the sum of power associated with each UE should be less than the maximum allowable power. Constrained $\textbf{C7}$ guarantees that the spectrum allocation vector of $n$th user is zero if not associated with $l$th UAV. Whereas $\textbf{C8}$ denotes that one sub-channel is allocated to only one user to prevent interference. Similarly, $\textbf{C9}$ states that the minimum percentage of UEs should be served. Similarly, $\textbf{C10}$ states that the distance between two UAVs should be greater than or equal to the minimum allowable distance. Whereas , $\textbf{C11}$ to $\textbf{C15}$ represents the lower and upper bounds of decision variables
\vspace{-1mm}
\section{Proposed Iterative Algorithm}
\label{Proposed Iterative Algorithm}
The optimization problem in \eqref{Main Equation1} is mixed-integer non-linear programming (MINLP) and NP-hard due to the integer nature of user association matrix $\!\!\textbf{J}$, UAVs placement vector $\!\!\Vec{I}\!\!$ and sub-carrier $\!\!\textbf{A}\!\!$ allocation matrix. Furthermore, it is challenging to get optimal results as it involves non-linear, non-convex objective functions and constraints. To find the optimal best solution, we need to calculate the minimum number of UAVs needed to provide communication services in a wireless communication network. Additionally, in OFDMA enabled UAV communication system, the maximum serving capacity of each UAV depends upon the maximum number of sub-carrier in a network. Therefore the maximum number of UE that each UAV can support is equal to the number of sub-carriers, i.e., $C^l_{max}\!\!=\!\!K$. Similarly, the number of estimated UAVs is calculated as $L\!\! =\!\! \left\lceil {\frac{ \lambda N}{{{C^{\max}}}}} \right\rceil$, Where $\lambda$ represents the percentage of users need to be served.
\par
Furthermore, the sum rate in \eqref{Rate carrier} depends upon the probability of line of sight and co-cell interference, which is the function of $3D$ location of the UAVs and transmission power, respectively. For the air-to-ground channel, UE's experience the line of sight communication with the probability represented by  $\!PLos_n^l$.  This  $\!PLos_n^l\!$ mainly depends on the angle of elevation $\!\left(\theta_n^l\right)$; as $\!\left(\theta_n^l\right)\!$ increases, $\!PLos_n^l$  increase, on the other hand, results in additional losses as the path loss between the UAVs and their associated UEs decreases\cite{7037248}. Thus, the desired objective can be achieved by deploying the UAVs at the point such that the probability of line-of-sight communication is greater than the threshold value $\phi$, i.e., $\!\!PLos_n^l \!\left(\theta_n^l\right)\! \ge\! \phi$ and user is only associated with that UAV located at a distance less than $\!{\scriptstyle\frac{{{h_l}}}{{\sin \left( {P{{_{nl}^{LoS}}^{^{ - 1}}}\left( \phi  \right)} \right)}}}$. Therefore, we decouple the original optimization problem into three subproblems, which are solved iteratively to find the final solutions. 

The UEs association problem, given the location of UAVs and UEs, can be formulated as follows:
\vspace{-1mm}
\begin{subequations}
\small
\vspace{-1mm}
	\label{Main Equation2a}
	\begin{align}
		\textbf{P2a}:&\; \mathop {\min }\limits_{\bf{J}} \sum\limits_{n = 1}^N {\sum\limits_{l = 1}^L {{J_{nl}}{{\left\| {{{\bf{V}}_n} - {{\bf{W}}_l}} \right\|}^2}} } \\[-1pt]
\textit{s.t.}~&~\text{\cref{C2,C3,C4,C9,C15}}\\[-2pt]
\vspace{-2mm}
	 &{J_{nl}}\left\| {{{\bf{V}}_n} \!\!- {{\bf{W}}_l}} \right\| \le \frac{{{h_l}}}{{\sin \left( {P{{_{nl}^{LoS}}^{^{ - 1}}}\left( \phi  \right)} \right)}},\forall n,l.
	\end{align}
	\vspace{-0.5mm}
\end{subequations}
In equation \eqref{Main Equation2a}, ${{\bf{V}}_n} \!\triangleq\!\left( {{x_n},{y_n},0} \right)$ represents the location of UEs. The optimization problem mentioned in \eqref{Main Equation2a} is linear programming (LP) and convex and can be solved easily using the Branch and Bound (BnB) algorithm.
\par
Given the user association matrix $\!\textbf{J}\!$ calculated in the user assignment step, the location of the UAV is updated such that the distance between the UAV and associated UEs is minimized mathematically can be expressed as follows:
\vspace{-1mm}
\begin{subequations}
\small
	\label{Main Equation2b}
	\begin{align}
	\label{P2b Objective}
		\textbf{P2b}:&\mathop {\min }\limits_{\bf{W}} \sum\limits_{n = 1}^{{Q_l}} {\left[ {{{\left( {{x_n} \!-\! {x_l}} \right)}^2} + {{\left( {{y_n}\! -\! {y_l}} \right)}^2} + h_l^2} \right],\forall l}  \\[-1pt]
		\label{P2b Constraint}
   \textbf{C1}:&{\left( {{x_n} \!\!-\!\! {x_l}} \right)^2}\!\! +\!\! {\left( {{y_n} - {y_l}} \right)^2}\!\! + h_l^2\xi  \le {\rm{0, }}\forall n \in Q_l\\[-1pt]
&\text{\cref{C10,C11,C12,C13}}.
		\end{align}
\end{subequations}
\vspace{-1mm}
In \eqref{Main Equation2b}, the quantity $Q_l$ denotes the number of UEs associated with the $l$-th UAV. Similarly, in \eqref{P2b Constraint}, the square of the distance between the UAV and associated UEs is influenced by ${\scriptstyle \xi \! = \!\left( {1 - \frac{1}{{\sin \left( {P_{nl}^{Lo{S^{ - 1}}}\left( \phi  \right)} \right)}}} \right)}$, ensuring that the UAV is positioned at a point where the angle elevation calculated based on its geographical location meets $\!PLos_n^l \left(\theta_n^l\right)\!\ge\!\!\phi$. The optimization problem in equation \eqref{Main Equation2b} is convex and can be efficiently solved using quadratic constraint quadratic programming (QCQP). To obtain the optimal $3D$ location of UAVs and their associated user matrix, the BnB and QCQP-based algorithms iteratively minimize the path loss between UAVs and their associated users, as detailed in Algorithm ~\ref{algo:Algorithem 1}.

\begin{algorithm2e}
\small
\SetAlgoLined
\textbf{Initialization: } $PL_{old}$, error\;
\textbf{Execution: }\;
\While{error $>$ $\epsilon$}{
        Given the random location of UAV's Solve \eqref{Main Equation2a} Using Branch and Bound (BnB) algorithm  to get the User association matrix $\textbf{J}$.\\[-1pt]
        Based upon the User Association matrix, update the location of UAV by solving \eqref{Main Equation2b}\\[-1pt]
        Compute Path Loss using using \eqref{channel} to obtain $PL_{*}$ \\
        Compute \textbf{error} = $|PL_{*}$-$PL_{old}|$\\[-1pt]
        Update $PL_{old}$=$PL_{old}-PL_{*}$
}
\caption{\textbf{\small{Iterative Algorithm to solve}} \eqref{Main Equation2a} \!\&\eqref{Main Equation2b}}
\label{algo:Algorithem 1}
\end{algorithm2e}
After determining the user association and UAV locations, we aim to minimize co-cell interference and maximize the sum rate required to meet the quality of service requirements through optimal sub-carrier allocation and transmit power. The third sub-problem is expressed mathematically as follows:
\vspace{-1mm}
\begin{subequations}
\vspace{-1mm}
	\label{Main Equation3}
	\begin{align}
	\textbf{P3a}:\;\;& \mathop {\max }\limits_\textbf{P, A} \sum\limits_{n = 1}^N {\sum\limits_{l = 1}^L R_n^l}\\[-1pt]
	\label{Main Equation3C1}
	\textbf{C1}:\;\;&{R_n^l \ge J_{nl}R_n},\forall n\in \mathcal{N}, l\in \mathcal{L} \\[-1pt]
	&\text{\cref{C5,C6,C7,C8,C14,C15}}.
		\end{align}
		\vspace{-1mm}
\end{subequations}
\vspace{-0.15mm}
The resource allocation problem in \eqref{Main Equation3} is non-convex and non-linear  in nature due to the rate function $R^l_n$ and constraint \eqref{Main Equation3C1}. To tackle this difficulty, an auxiliary variables $\eta_n^l, s_n^{k,l}, \forall n,l$ are introduced and reformulate \eqref{Main Equation3} as 
\vspace{-1mm}
\begin{subequations}
\small
	\label{Transformed2}
	\begin{align}
	\label{Convex Equation Objective 2}
	\textbf{P3b}:&\;\mathop {\max }\limits_{{\bf{A}},{\bf{P}},{\bf{\eta }},{\mathfrak{s}}} \sum\limits_{n \in N} {\sum\limits_{l \in L} {\eta _n^l} } \\[-3pt]
	\begin{split}
	\label{T2C1}
	\textbf{C1}:&\;\sum\limits_{k \in \mathcal{K}} \log \left(\Phi_n^{k,l}+\delta^2+p_n^{k,l}g_n^{k,l}\right) \ge \frac{{\log (2)\eta _n^l}}{B}\\[-5pt]
	&\qquad+ \sum\limits_{k \in \mathcal{K}}s_n^{k,l}, \forall n \in \mathcal{N}, l \in \mathcal{L} 
	\end{split}\\[-5pt]
	\label{C3T3}
	\textbf{C2}:&\;\Phi_n^{k,l}+\delta^2 \le e^{s_n^{k,l}}, \forall n\in \mathcal{N},l\in \mathcal{L},k\in \mathcal{K}\\[-3pt]
	\textbf{C3}:&\;\eta^l_n \ge J^l_n R^l_n, \forall n \in \mathcal{N}, l \in \mathcal{L} \label{C3T2} \\[-3pt]
	&\text{\cref{C5,C6,C7,C8,C14,C15}},
	\end{align}
\end{subequations}
\vspace{-0.5mm}
where \eqref{C3T2} guarantee the QoS requirement, and \eqref{T2C1} and \eqref{C3T3} equivalently represent $R_n^l \geq \eta_n^l$.  
Solving \eqref{Transformed2} is still challenging due to the non-convex constraint \eqref{C3T3}. Fortunately, it has a form of different-of-convex (DC) constraint; hence we can employ the first-order approximation to convexify the right-hand-side of \eqref{C3T3}, which can be reformulated as follows:
\vspace{-0.5mm}
\begin{subequations}

	\label{Transformed3}
	\begin{align}
	\label{Convex Equation Objective 3}
	\textbf{P3c}:&\;\mathop {\max }\limits_{{\bf{A}},{\bf{P}},{\bf{\eta }},{\mathfrak{s}}} \sum\limits_{n \in N} {\sum\limits_{l \in L} {\eta _n^l} } \\
	\label{C3T4}
	\textbf{C1}:&\;{\Phi_n^{k,l}+\delta^2\le e^{{s_o}_n^{k,l}}\left(s_n^{k,l}-{s_o}_n^{k,l}+1\right)},\notag \\
	&\qquad\qquad \qquad\qquad\forall n \in \mathcal{N},k\in\mathcal{K},l\in\mathcal{L} \\
	&\text{\cref{T2C1,C3T2,C5,C6,C7,C8,C14,C15}},
	\end{align}
\end{subequations}
\vspace{-0.5mm}
Given a feasible point ${s_o}_n^{k,l}$ for constraint \eqref{C3T3}, the objective function being linear and all constraints being convex, problem \eqref{Transformed3} is a convex optimization problem. Therefore, it can be efficiently solved using standard methods such as the interior-point method. However, the solution of \eqref{Transformed3} is dependent on the feasible point ${s_o}$. To tackle this, we propose an iterative algorithm (Algorithm 2), which involves a sequence of convex optimizations to approximate the best solution for problem \eqref{Transformed2}. By iteratively updating the feasible point, our algorithm effectively converges to the optimal solution of the problem.
\begin{algorithm2e}
\small
\SetAlgoLined
\textbf{Initialization: } $s_o$, $\eta_{old}$, error\;
\textbf{Execution: }\;
\While{error $>$ $\epsilon$}{
        Given the location of UAVs and user association matrix form Algorithm \ref{algo:Algorithem 1}, Solve \eqref{Transformed3} to get the optimal value of $\eta _*$,$a _*$, $p_*$,$s_*$ \\
        Compute \textbf{error} = $|\eta_{*}$-$\eta_{old}|$\\
        Update $\eta_{old}$=$\eta_{*}$,$s_{o}$=$s_{*}$
}
\caption{\textbf{{Iterative Algorithm to solve} \eqref{Transformed2}}}\label{algo:Algorithem 2}
\end{algorithm2e}
\vspace{-1mm}
\section{Results and Discussion}
\label{Results and Discussion}
\vspace{-1mm}
This section demonstrates the effectiveness of the proposed joint algorithm via extensive Monte-Carlo simulations. In our simulations, we assume the $N$ number of users distributed randomly over an $\!1\!\!\!$ km $\!\!\times\!\!\!$  $1\!\!\!$ km area. Furthermore, $L$ number of UAVs are deployed to provide communication services using the OFDMA protocol. Whereas other simulation parameters are as follows: $f_c\!\!=\!\!\!1$ GHz, $\alpha\!\!=\!\!4$, $K\!\!=\!\!8$ and $\delta^2\!\!=\!\!-10$ dBm. Moreover, the effectiveness of the proposed scheme is illustrated by considering the users' path loss and sum rate as a performance matrix for Algorithms \ref{algo:Algorithem 1} and  \ref{algo:Algorithem 2}, respectively.

\begin{figure}
	\centering
	\includegraphics[width=1\linewidth]{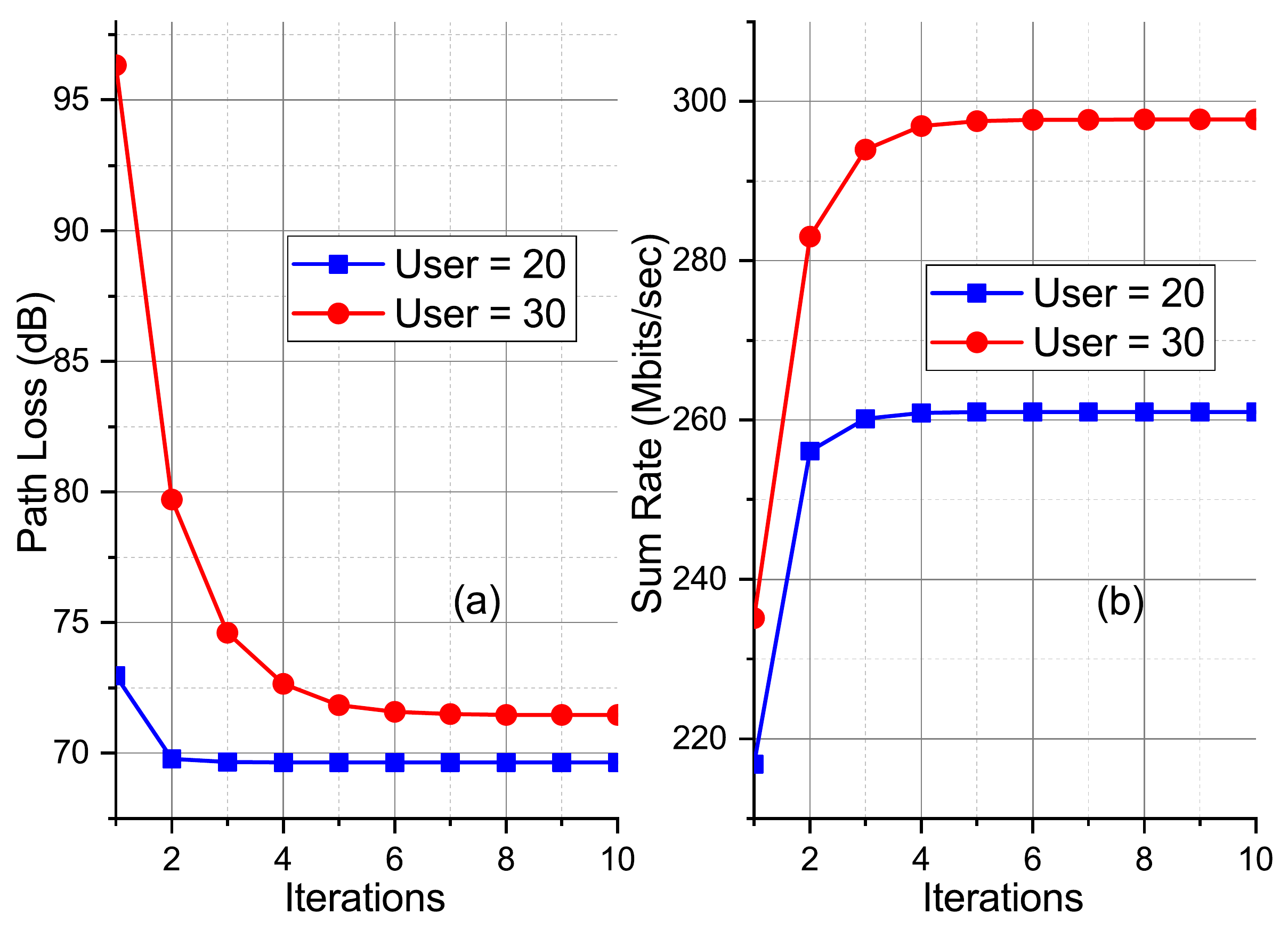}
	\caption{Convergence analysis}
	\label{fig:Convergence analysis}
	\vspace{-7mm}
\end{figure}
\par
Furthermore, the convergence of the proposed algorithms is compared with different stimulation parameters, as shown in Fig. \ref{fig:Convergence analysis}. Fig.\ref{fig:Convergence analysis}a represents the convergence graph of Algorithm \ref{algo:Algorithem 1} by considering path loss as a performance matrix. Results reveal that as the number of iterations increases, the path loss of the network also decreases. This trend is because UAVs update their locations and user association matrix incrementally in the number of iterations. As a result, the distance between  UAVs and associated users decreases; hence path loss decreases.
Similarly, based on the UAV's locations and user association matrix, we solve the sum rate maximization problem using Algorithm \ref{algo:Algorithem 2}. Results in Fig.~\ref{fig:Convergence analysis}b demonstrate the efficiency of the proposed iterative algorithm.
\begin{figure}[]
	\begin{subfigure}{0.9\columnwidth}
		\centering
		\includegraphics[width=0.8\linewidth]{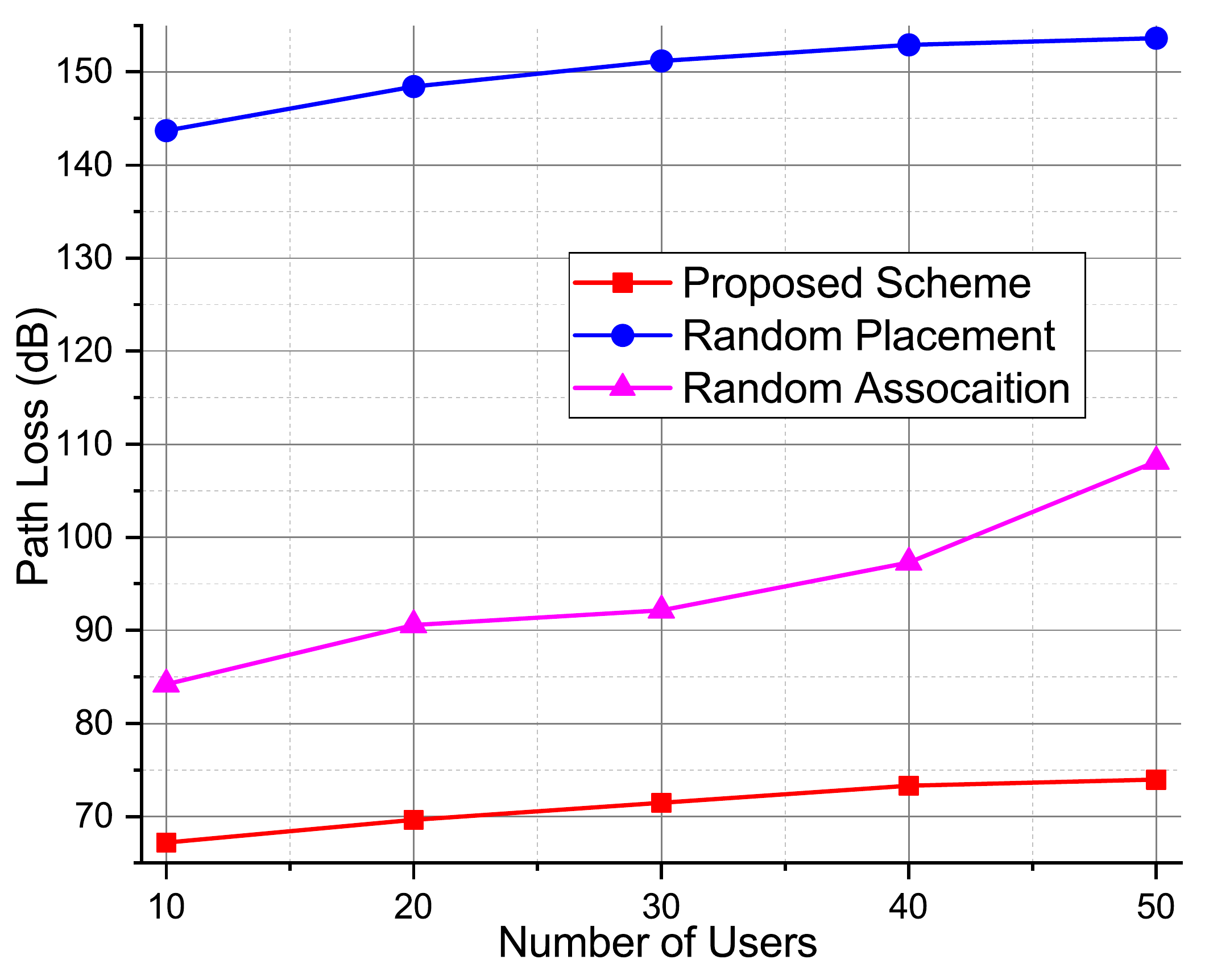}
		\caption{Path Loss vs Number of users}
		\label{fig:PathLoss_D_Cases}
	\end{subfigure} 
	\hfill
	\begin{subfigure}{0.9\columnwidth}
		\centering
		\includegraphics[width=0.8\linewidth]{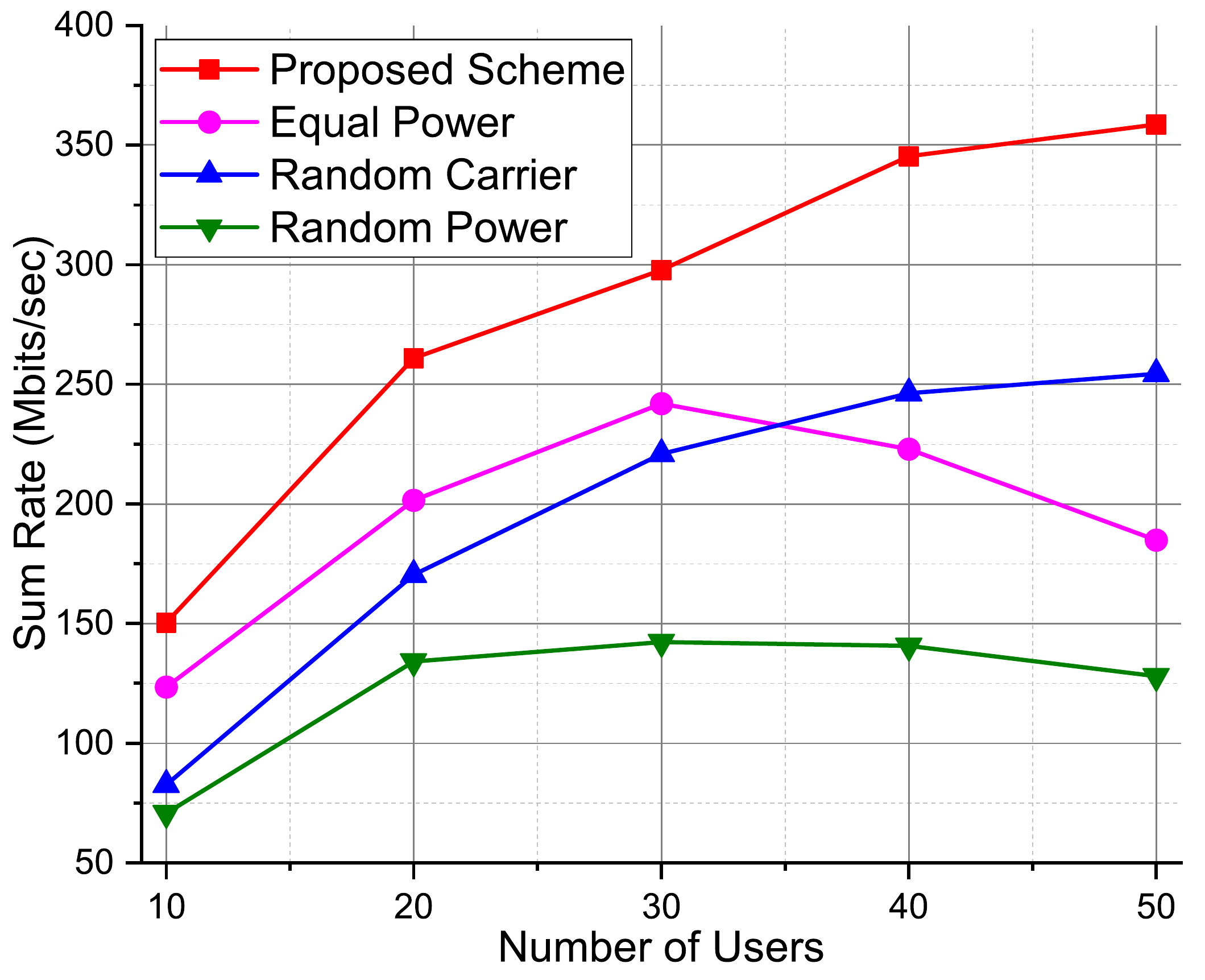}
		\caption{Sum Rate vs Number of users}
		\label{fig:SumRate_D_Cases}
	\end{subfigure}
	\caption{Comparison of Proposed scheme with some baseline schemes}
	\vspace{-7mm}
	\label{fig: Comparison}
\end{figure}
\par

After validating the convergence behaviour of Algorithms \ref{algo:Algorithem 1} and \ref{algo:Algorithem 2}, the performance of the proposed solution is exhibited by considering sum rate (Mbps) as the performance metric. The sum rate of the UEs mainly depends on the path loss, which is the function of $3D$ coordinates of the UAV and user association matrix. Accordingly, the optimal deployment of UAVs and optimal allocation of users results in the minimization of the path loss; consequently, the sum rate of the user increases. Besides, the optimal deployment of UAVs will result in the minimization of transmission power. Therefore, to verify the effectiveness of the proposed scheme, we compare the proposed design with baseline schemes named: random deployment of UAV with optimal allocation of users and optimal deployment of UAV with random allocation of users , as shown in Fig.~\ref{fig: Comparison}. Fig. \ref{fig:PathLoss_D_Cases} represents the performance analysis of the system by considering path loss as a performance metric across the number of users. It is shown that the proposed scheme achieves the minimum path loss compared to random deployement and random association solutions. As a result, the transmission power of the UAVs decreases, which leads to minimizing co-cell interference or helps to increase the sum rate of users. Whereas, based on the optimal location of UAV and user association matrix, the performance is further evaluated by solving optimization problem \eqref{Transformed3} using Algorithm \ref{algo:Algorithem 2}.  Furthermore, the results of the proposed scheme is compared with equal/random power allocation and random carrier allocation scheme. Fig. \ref{fig:SumRate_D_Cases} reveals that optimal power allocation efficiently mitigates co-cell interference, resulting in a higher sum rate. 
\begin{figure}[]
	\begin{subfigure}{.9\columnwidth}
	\centering
	\includegraphics[width=0.8\linewidth]{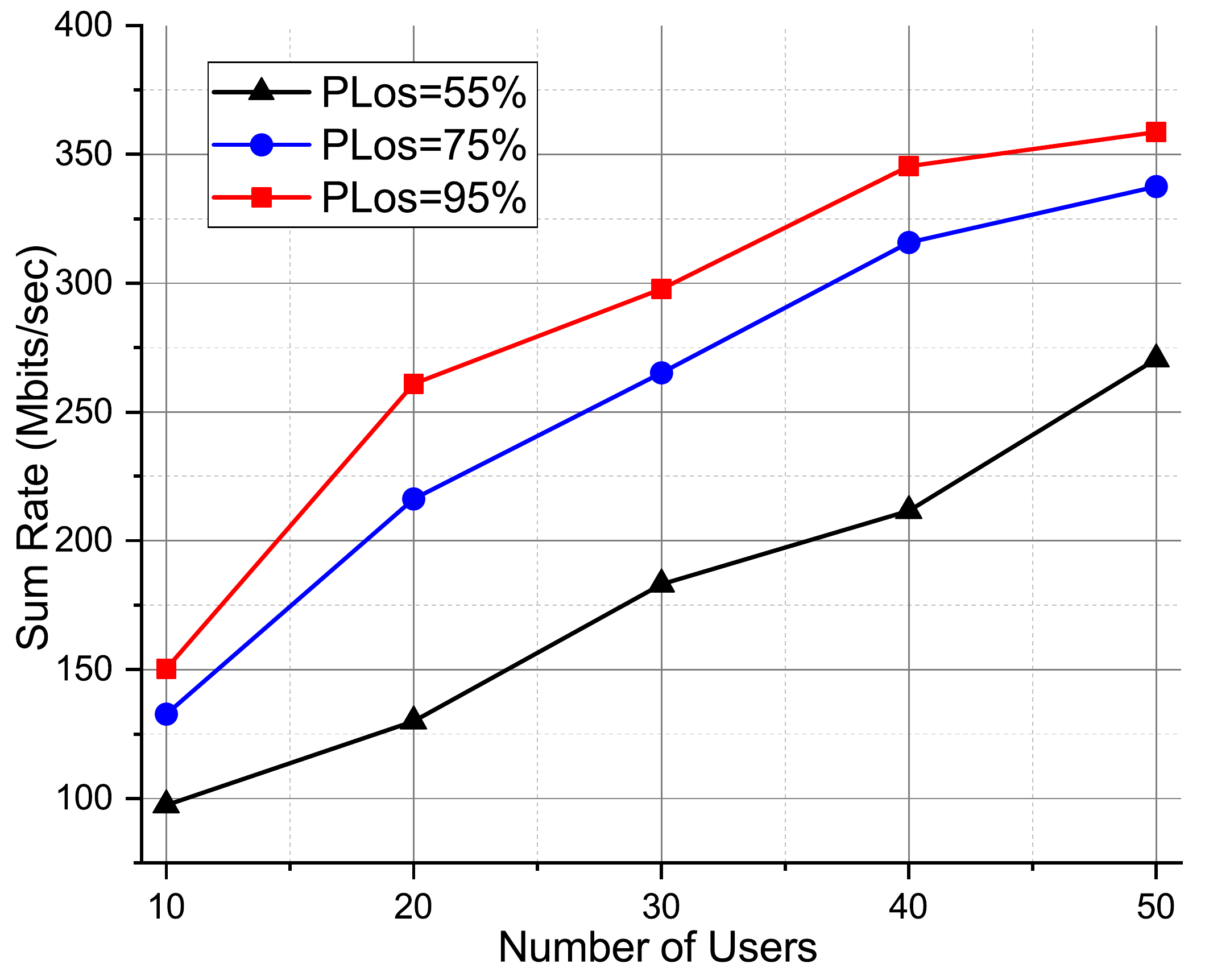}
	\caption{Sum Rate vs Number of Users}
	\label{fig:Figure_3_SumRate_vs_number_of_Usres}
	\end{subfigure} 
	\hfill
	\begin{subfigure}{.9\columnwidth}
		\centering
		\includegraphics[width=0.8\linewidth]{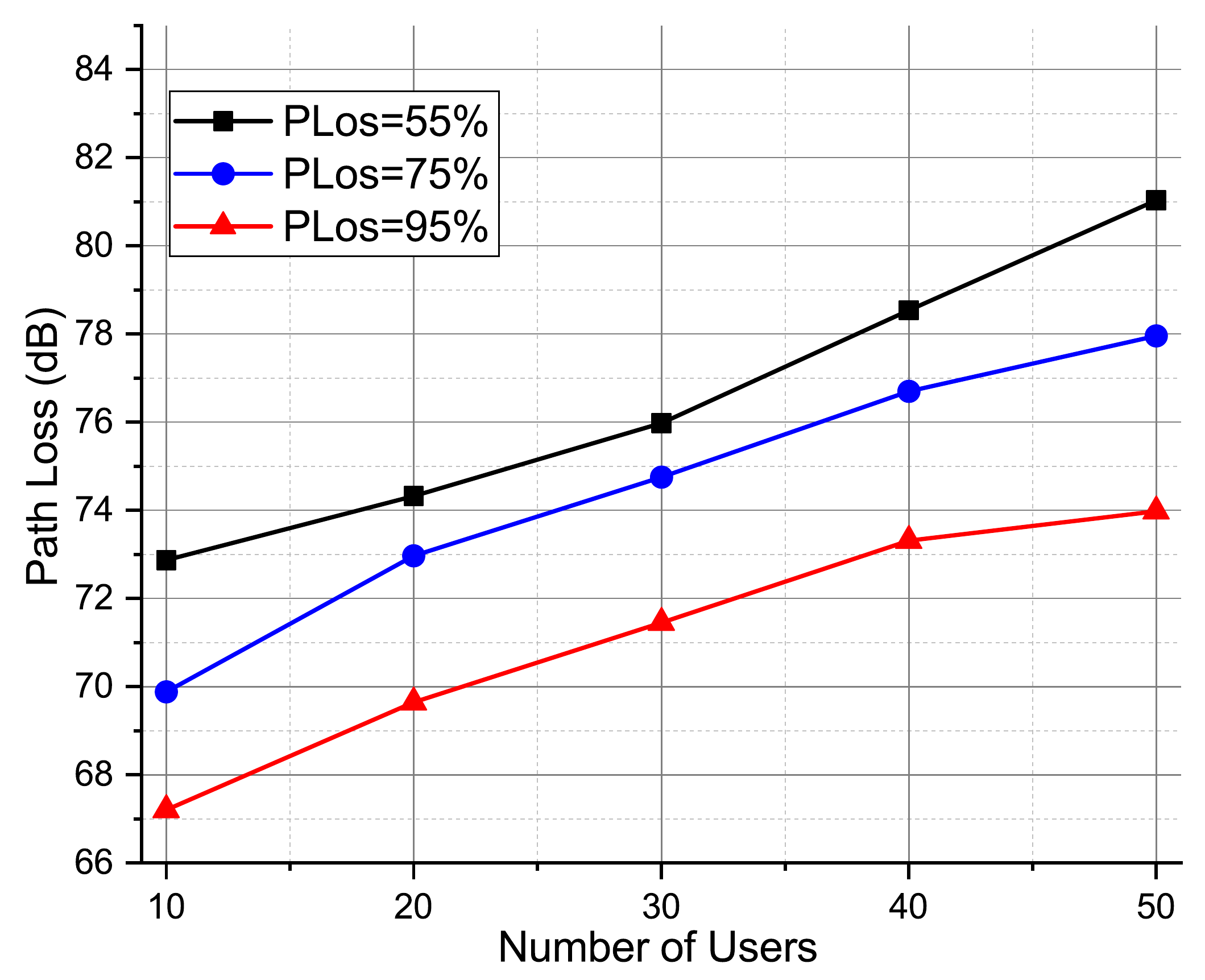}
		\caption{Path Loss vs Number of Users}
		\label{fig:Figure_3_PathLoss_vs_number_of_Usres}
	\end{subfigure}
	\caption{Comparison of Proposed scheme with the probability of line of sight}
	\vspace{-7mm}
	\label{fig: Probility}
\end{figure}

\par
In Fig.~\ref{fig: Probility}, we evaluate the impact of the probability of line of sight between the UAVs and their associated users on the system performance. It is shown in Fig.~\ref{fig:Figure_3_SumRate_vs_number_of_Usres} that increasing the PLoS will result in a higher system sum rate. This is because as the PLoS increases from $55\%$ to $95\%$, path loss between the UAVs and associated users decreases significantly, as shown in Fig. \ref{fig:Figure_3_PathLoss_vs_number_of_Usres}, resulting in the higher system sum rate. 
\vspace{-1.5mm}
\section{Conclusion}
\label{Conclusion}
\vspace{-1.5mm}
In this work, we have considered that the ODFMA enables multiple UAV communication systems to provide communication services to users in remote areas. We proposed a collaborative framework for determining the optimal UAV deployment to minimize the UAV's transmission power while still meeting the users' QoS requirements. Path loss and user sum rate are performance metrics used to assess the proposed system's performance. The results revealed that optimal deployment and user association minimize path loss, which directly impacts the sum rate of users. Additionally, the network's performance is evaluated across multiple PLoS. The results demonstrate that by increasing the PLoS, the path loss decreases significantly, thereby increasing the user's sum rate.
\vspace{-0.5mm}
\section{Acknowledgement}
\vspace{-0.5mm}
This work was supported by the Luxembourg National Research Fund via project 5G-Sky, ref. FNR/C19/IS/13713801/5G-Sky, and project RUTINE, ref.FNR/C22/IS/17220888/RUTINE.
\vspace{-1.5mm}
\balance
\vspace{-2mm}
\small
\bibliographystyle{IEEEtran}
\bibliography{ReferenceBibFile}
\end{document}